\documentclass[12pt,a4paper]{article}
\usepackage{graphicx}
\usepackage[utf8]{inputenc}
\usepackage[T1]{fontenc}
\usepackage{amssymb,amsmath,amsfonts}
\usepackage{bbm}
\usepackage{braket}

\usepackage[font=footnotesize]{caption}

\newcommand{\up}{\uparrow}
\newcommand{\down}{\downarrow}
\newcommand{\Lcal}{\mathcal{L}}
\newcommand{\Tr}{\mathrm{Tr}}
\usepackage[breaklinks=true,colorlinks=true,linkcolor=blue,urlcolor=blue,citecolor=blue]{hyperref}

\title{On the spreading of quantum walks starting from local and delocalized states}

\author{Alexandre C. Orthey Jr.\footnote{Departamento de F\'isica, Universidade do Estado de Santa Catarina, 89219-710, Joinville, SC, Brazil.} \quad\&\quad Edgard P. M. Amorim$^*$\footnote{Corresponding author: edgard.amorim@udesc.br}}

\begin{document}

\maketitle

\begin{abstract}
We investigate the ballistic spreading behavior of the one-dimensional discrete time quantum walks whose time evolution is driven by any balanced quantum coin. We obtain closed-form expressions for the long-time variance of position of quantum walks starting from any initial qubit (spin-$1/2$ particle) and position states following a delta-like (local), Gaussian and uniform probability distributions. By averaging over all spin states, we find out that the average variance of a quantum walk starting from a local state is independent of the quantum coin, while from Gaussian and uniform states it depends on the sum of relative phases between spin states given by the quantum coin, being non-dispersive for a Fourier walk and large initial dispersion. We also perform numerical simulations of the average probability distribution and variance along the time to compare them with our analytical results.
\end{abstract}

\vspace{2pc}

\section{Introduction}

The quantum counterparts of the classical random walks are known as quantum random walks \cite{aharonov1993quantum,kempe2003quantum} or quantum walks. The quantum walker is a qubit, a particle with an internal degree of freedom (spin $1/2$-like) placed on a regular lattice where each site is an external degree of freedom (position). Instead of a coin tossing game, to determine whether the particle goes to left or right, the dynamics is given by a unitary operator applied successive times to an initial state time-evolving the quantum state. This operator is formed by a quantum coin which rotates the qubit followed by a conditional displacement that displaces the qubit according to its spin state. The main difference between classical and quantum walks is due to the superposition principle, which leads the quantum walks to have unique features: a double peak distribution with a quadratic gain in their variance of the position along the time and the entanglement between the internal and external degrees of freedom created by their particular dynamics \cite{venegas2012quantum}.  

Quantum walks have been attracting a lot of attention due to their diversity of the implications in basic science and their potential applications. For instance, they are useful for the understanding of some biological processes such as the photosynthesis \cite{engel2007evidence} or human decision-making \cite{buseymer2006quantum}, to perform computational tasks as quantum search engine \cite{shenvi2003quantum,tulsi2008faster}, make universal quantum computation \cite{childs2009universal,lovett2010universal}, for generating maximal entanglement \cite{vieira2013dynamically} and quantum localization \cite{vieira2014entangling}. Moreover, they are versatile enough to be implemented in some experimental platforms \cite{wang2013physical}. 

The main purpose here is to understand how the delocalization of the initial state affects the spreading behavior of the quantum walks. Few earlier works discuss some aspects of this issue \cite{tregenna2003controlling,brun2003quantum,valcarcel2010tailoring,zhang2016creating}, however, a general answer for all possibilities of initial spin states (qubits) or for any type of quantum coin is still missing. To achieve this aim, we use a mathematical framework to obtain via analytical \cite{brun2003quantum} and numerical approaches the variance of the position regarding local, Gaussian and uniform states for any initial qubit and balanced coin. In particular, we also calculate the average quantities to analyze the general features and differences among these states.

The article is structured as follows. In Section \ref{sec:2}, we review the mathematical formalism of quantum walks, their initial states and dynamical evolution. In Section \ref{sec:3}, we obtain a general expression of the long-time variance of the position in the momentum space and calculate analytical expressions for the variance of quantum walks starting from local, Gaussian and uniform states. We also discuss the dispersion velocity as function of the initial spin state for local and uniform states. In Section \ref{sec:4}, we confront the average variance calculated from our models to the numerical simulations. Finally, a brief conclusion with our main results is depicted in Section \ref{sec:5}.  

\section{One-dimensional quantum walks} \label{sec:2}

Formally, a quantum walk state belongs to the Hilbert space $\mathcal{H}=\mathcal{H}_C\otimes\mathcal{H}_P$, where $\mathcal{H}_C$ is the coin space and $\mathcal{H}_P$ is the position space. The coin space is a complex two-dimensional vector space spanned by two spin states $\{\ket{\up}, \ket{\down}\}$ and the position space is a countable infinite-dimensional vector space spanned by a set of orthonormal vectors $\{\cdots,\ket{j-1},\ket{j},\ket{j+1},\cdots\}$ with $j\in\mathbb{Z}$ being the discrete positions on a lattice. The one-dimensional quantum walker is a qubit, a particle with an internal degree of freedom as a two-level state (spin $1/2$-like) and its position and momentum as the external degrees of freedom. Then, let us first consider an arbitrary initial qubit state,
\begin{equation}
\ket{\Psi_s(0)}=\cos\left(\frac{\alpha}{2}\right)\ket{\up}+e^{i\beta}\sin\left(\frac{\alpha}{2}\right)\ket{\down},
\label{Psi_s}
\end{equation}
in the Bloch sphere representation \cite{nielsen2010quantum} where $\alpha \in[0,\pi]$ and $\beta \in[-\pi,\pi]$. For instance, an up spin state has $\alpha=0$ for any value of $\beta$ or still, an equal superposition of spin states without phase difference between them has $\alpha=\pi/2$ and $\beta=0$. Therefore, a general quantum walk state is 
\begin{equation}
\ket{\Psi(0)}=\sum_{j=-\infty}^{+\infty}\ket{\Psi_s(0)}\otimes f(j)\ket{j},
\label{Psi_0}
\end{equation}
where $a(j,0)=f(j)\cos(\alpha/2)$ and $b(j,0)=f(j)e^{i\beta}\sin(\alpha/2)$ are the initial spin up and down amplitudes, respectively, and $|f(j)|^2$ gives us an initial probability distribution function. The condition of normalization is $\sum_j[|a(j,0)|^2+|b(j,0)|^2]=1$ with the sum over all integers.

We employ here a local, Gaussian and uniform initial states. Since the local state has a Dirac delta function $\delta(j)$ as the initial probability distribution, we obtain a qubit on the origin position, 
\begin{equation}
\ket{\Psi_L(0)}=\ket{\Psi_s(0)}\otimes\ket{0}.
\label{Psi_0_Local}
\end{equation}
Taking a Gaussian probability distribution with initial dispersion $\sigma_0$, then a general initial Gaussian state is
\begin{equation}
\ket{\Psi_G(0)}=\sum_{j=-\infty}^{+\infty}\ket{\Psi_s(0)}\otimes\frac{\text{exp}\left(-j^2/4\sigma_0^2\right)}{(2\pi\sigma_0^2)^{\frac{1}{4}}}\ket{j},
\label{Psi_0_Gauss}
\end{equation}
while a uniform state could be written as
\begin{equation}
\ket{\Psi_U(0)}=\sum_{j=-\infty}^{+\infty}\ket{\Psi_s(0)}\otimes u\ket{j},
\label{Psi_0_Unif}
\end{equation}
with $\sum_j|u|^2=1$ and $u\rightarrow 0$ being the sum over all integers.

The unitary dynamical evolution of the quantum walk starting from an initial state $\ket{\Psi(0)}$ is given by,
\begin{equation}
\ket{\Psi(t)}=U(q,\theta,\phi)^t\ket{\Psi(0)},
\label{time_evolution}
\end{equation}
in discrete time steps $t$ with
\begin{equation}
U(q,\theta,\phi)=S.[C(q,\theta,\phi)\otimes\mathbbm{1}_P],
\label{U_operator}
\end{equation}
being the time evolution operator where $\mathbbm{1}_P$ is the identity operator in $\mathcal{H}_P$, $C(q,\theta,\phi)$ is the quantum coin and $S$ is the conditional displacement operator.

The quantum coin $C(q,\theta,\phi)$ operates over the spin states and generates a superposition of them. Since a general quantum coin $C(q,\theta,\phi)$ belongs to the $SU(2)$ and up to an irrelevant global phase, it can be written in the following way,
\begin{equation}
\displaystyle
C(q,\theta,\phi) = 
\begin{pmatrix}
\sqrt{q} & \sqrt{1-q}e^{i\theta} \\
\sqrt{1-q}e^{i\phi} & -\sqrt{q}e^{i(\theta+\phi)}
\end{pmatrix},
\label{Quantum_coin}
\end{equation}
with three independent parameters $q$, $\theta$ and $\phi$. The first parameter ranges from $0$ to $1$, and it determines if the coin is unbiased ($q=1/2$) or biased ($q\neq1/2$). The last both terms range from $0$ to $2\pi$, and they control the relative phases between spin states.

The conditional displacement operator $S$,
\begin{equation}
S=\sum_j(\ket{\up}\bra{\up}\otimes\ket{j+1}\bra{j}+\ket{\down}\bra{\down}\otimes\ket{j-1}\bra{j}),
\label{S_operator}
\end{equation}
shifts the qubit from the site $j$ to the site $j+1$ ($j-1$) conditioned to its up (down) spin state, which generates entanglement between the spin and position states \cite{vieira2013dynamically,vieira2014entangling}. 

The one time step evolution of a quantum walk state gives
\begin{equation}
\ket{\Psi(t)}=U(q,\theta,\phi)\ket{\Psi(t-1)}
\label{onestep_Psi}
\end{equation}
and by replacing \eqref{Quantum_coin} and \eqref{S_operator} into the $U(q,\theta,\phi)$ in \eqref{onestep_Psi} allow us to write the following equations for the spin amplitudes $a(j,t)$ and $b(j,t)$,
\begin{align}
a(j,t)&=\sqrt{q}a(j-1,t-1)+\sqrt{1-q}e^{i\theta}b(j-1,t-1),\\
b(j,t)&=\sqrt{1-q}e^{i\phi}a(j+1,t-1)-\sqrt{q}e^{i(\theta+\phi)}b(j+1,t-1).
\end{align}
These recurrence relations are used to obtain the quantum walk probability distribution profile and their variance of position through an iterative calculation starting from their initial amplitudes $a(j,0)$ and $b(j,0)$. This approach is used to perform the numerical calculations. The probability distribution for each position $j$ at a time step $t$ can be obtained straightforwardly as 
\begin{equation}
|\Psi(j,t)|^2=|a(j,t)|^2+|b(j,t)|^2,
\label{Prob_Dist}
\end{equation}
such that $|a(j,t)|^2$ and $|b(j,t)|^2$ are the probability distributions for up and down spin components. The variance of position at a particular time step $t$ is
\begin{equation}
\sigma^2(t)=\braket{\hat{\mathbf{j}}^2}_t-\braket{\hat{\mathbf{j}}}_t^2,
\label{variance}
\end{equation}
where
\begin{equation}
\braket{\hat{\mathbf{j}}^m}_t=\sum_j j^m|\Psi(j,t)|^2.
\label{variance_sup}
\end{equation}

In order to deal with \eqref{variance_sup}, on the next section we will use a mathematical framework to develop closed-form expressions for the long-time variance of a quantum walk starting from any qubit on one position (local state) or spread over many positions following a Gaussian and uniform probability distributions.

\section{Long-time variance} \label{sec:3}

Since the expression \eqref{variance_sup} is difficult to handle analytically, we should made a change of basis to the dual $k$-space $\mathcal{\tilde{H}}_k$ spanned by the Fourier transformed vectors $\ket{k}=\sum_{j}e^{ikj}\ket{j}$ with $k\in[-\pi,\pi]$ \cite{ambainis2001one}. Then, the initial state \eqref{Psi_0} can be rewritten as,
\begin{equation}
\ket{\tilde{\Psi}(0)}=\int_{-\pi}^{\pi} \dfrac{\mathrm{d}k}{2\pi}\ket{\Phi_k(0)}\otimes\ket{k},
\label{Psitil_0}
\end{equation}
where $\ket{\Phi_k(0)}=\tilde{a}_k(0)\ket{\up}+\tilde{b}_k(0)\ket{\down}$. In the Fourier representation, the conditional displacement operator $S$ is diagonal,
\begin{equation}
S_k=\ket{\up}\bra{\up}\otimes e^{-ik}\ket{k}\bra{k}+ \ket{\down}\bra{\down}\otimes e^{ik}\ket{k}\bra{k},
\label{S_diag}
\end{equation} 
thus the time evolution operator \eqref{U_operator} in the $k$-space gives
\begin{equation}
U_k=\dfrac{1}{\sqrt{2}}\begin{pmatrix}
e^{-i(\delta+k)} & e^{i(\eta-k)} \\
e^{-i(\eta+k)} & -e^{i(\delta+k)}
\end{pmatrix},
\label{Uk}
\end{equation}
where $\delta=(\theta+\phi)/2$, $\eta=(\theta-\phi)/2$ and the coin is balanced ($q=1/2$). After the diagonalization of \eqref{Uk} we obtain the following eigenvalues and eigenvectors \cite{tregenna2003controlling},
\begin{equation}
\lambda_k^{\pm}=\pm\dfrac{e^{i\delta}}{\sqrt{2}}\left[\sqrt{1+\cos^2(k-\delta)}\mp i\sin(k-\delta)\right],\\
\label{eigenvalues}
\end{equation}
\begin{align}
\ket{\Phi_k^{\pm}}&=\dfrac{1}{N_k^{\pm}}
\begin{pmatrix}
e^{ik} \\
e^{-i(\delta+\eta)}\left(\sqrt{2}\lambda_k^{\pm}-e^{ik}\right)
\end{pmatrix},
\label{eigenvectors}
\end{align}
with
\begin{equation}
(N_k^{\pm})^2=4\mp 2\left[\cos (k-\delta)\sqrt{1+\cos^2 (k-\delta)}\pm \sin^2 (k-\delta)\right].
\label{N_k}
\end{equation}
Now we are able to elaborate some details of the formalism introduced by Brun \textit{et al.} \cite{brun2003quantum} in order to achieve an analytical expression for the variance of the position without disorder, nor decoherence. The density operator can be written from \eqref{Psitil_0} as,
\begin{equation}\label{densidade}
\rho(0)=\ket{\tilde{\Psi}(0)}\bra{\tilde{\Psi}(0)}=\int_{-\pi}^{\pi}\dfrac{\mathrm{d}k}{2\pi}\int_{-\pi}^{\pi}\dfrac{\mathrm{d}k'}{2\pi}\ket{\Phi_k(0)}\bra{\Phi_k(0)}\otimes\ket{k}\bra{k'}.
\end{equation}
Let us rewrite the operator $U_k$ from  \eqref{Uk} in the following way,
\begin{equation}\label{U_k_decomposto}
U_k=\left(e^{-ik}\ket{\up}\bra{\up}+e^{ik}\ket{\down}\bra{\down}\right)C(q,\theta,\phi).
\end{equation}
Since an arbitrary operator $\Lambda$ is transformed as
\begin{equation}
\mathcal{L}_{kk'}\Lambda= U_k\Lambda  U_{k'}^\dagger,
\end{equation}
thus, by using this notation $\rho_k(0)=\ket{\Phi_k(0)}\bra{\Phi_k(0)}$, we can write
\begin{equation}
\rho(t)=\ket{\tilde{\Psi}(t)}\bra{\tilde{\Psi}(t)}=\int_{-\pi}^{\pi}\dfrac{\mathrm{d}k}{2\pi}\int_{-\pi}^{\pi}\dfrac{\mathrm{d}k'}{2\pi}\mathcal{L}_{kk'}^t\rho_k(0)\otimes\ket{k}\bra{k'}.
\end{equation}
The reduced density operator relative to the position is given by
\begin{equation}
\rho_P(t)=\int_{-\pi}^{\pi}\dfrac{\mathrm{d} k}{2\pi}\int_{-\pi}^{\pi}\dfrac{\mathrm{d} k'}{2\pi}\ket{k}\bra{k'}\Tr\left\{\mathcal{L}_{kk'}^t\rho_k(0)\right\},
\end{equation}
therefore, the probability to find the quantum walker on the position $j$ at time $t$ is
\begin{align}
|\Psi(j,t)|^2 &= \Tr\left\{ \left(\mathbbm{1}_C \otimes \ket{j}\bra{j}\right) \rho(t)\right\}, \\
              &= \Tr\left\{\ket{j}\bra{j} \rho_P(t)\right\}, \\
              &= \braket{j|\rho_P(t)|j}.
\end{align}
Since the inverse Fourier transform is
\begin{equation}
\ket{j}=\int_{-\pi}^{\pi}\dfrac{\mathrm{d}k}{2\pi}e^{-ikj}\ket{k},
\end{equation}
we have,
\begin{align}
|\Psi(j,t)|^2&=\int_{-\pi}^{\pi}\dfrac{\mathrm{d}k}{2\pi}\int_{-\pi}^{\pi}\dfrac{\mathrm{d}k'}{2\pi}\braket{j|k}\braket{k'|j}\text{Tr}\{\mathcal{L}_{kk'}^t\rho_k(0)\}\\
      &=\int_{-\pi}^{\pi}\dfrac{\mathrm{d}k}{2\pi}\int_{-\pi}^{\pi}\dfrac{\mathrm{d}k'}{2\pi}e^{-ij(k'-k)}\text{Tr}\{\mathcal{L}_{kk'}^t\rho_k(0)\}.
\end{align}
We calculate the expressions for the moments of the distribution by,
\begin{align}
\braket{\hat{\mathbf{j}}^m}_t &= \sum_j j^m |\Psi(j,t)|^2\nonumber \\
\label{jm}           &= \sum_j j^m \int_{-\pi}^{\pi}\dfrac{\mathrm{d}k}{2\pi} \int_{-\pi}^{\pi}\dfrac{\mathrm{d}k'}{2\pi} e^{-ij(k'-k)} \text{Tr}\{\mathcal{L}_{kk'}^t\rho_k(0)\},
\end{align}
where the position operator $\hat{\mathbf{j}}$ acts like $\hat{\mathbf{j}}\ket{j}=j\ket{j}$. We can change the order of sum and integration in  \eqref{jm} to use the identity,
\begin{equation}
2\pi(-i)^m\delta^{(m)}(k'-k)=\sum_{j=-\infty}^{+\infty}j^me^{-ij(k'-k)},
\end{equation}
where $\delta^{(m)}(k'-k)$ is the $m$-th derivative of Dirac delta function. Thus,
\begin{equation}
\braket{\hat{\mathbf{j}}^m}_t =\dfrac{(-i)^m}{2\pi}\int_{-\pi}^{\pi}\mathrm{d}k \int_{-\pi}^{\pi} \mathrm{d}k' \delta^{(m)}(k'-k) \text{Tr}\{\mathcal{L}_{kk'}^t\rho_k(0)\}.\label{jm_t}
\end{equation}
The integration above will be made by parts using the derivative of $\Lcal_{kk'}$ in terms of \eqref{U_k_decomposto},
\begin{align}
\dfrac{\partial}{\partial k} U_k\Lambda U_{k'}^\dagger &= \left(-ie^{-ik}\ket{\up}\bra{\up}+ie^{ik}\ket{\down}\bra{\down}\right) C(q,\theta,\phi)\Lambda U_{k'}^\dagger\nonumber\\
&=-i\left(e^{-ik}\ket{\up}\bra{\up}-e^{ik}\ket{\down}\bra{\down}\right) C(q,\theta,\phi)\Lambda U_{k'}^\dagger\nonumber\\
&=-i\left(\ket{\up}\bra{\up}-\ket{\down}\bra{\down}\right) U_k\Lambda U_{k'}^\dagger\nonumber\\
&=-i\hat{Z} U_k\Lambda U_{k'}^\dagger,
\end{align}
where $\hat{Z}=\ket{\up}\bra{\up}-\ket{\down}\bra{\down}$. Trace properties give us,
\begin{align}
    \dfrac{\partial}{\partial k}\Tr \{\Lcal_{kk'}\rho_k(0)\} &=\Tr \left\{\dfrac{\partial}{\partial k}\Lcal_{kk'}\rho_k(0)\right\}\nonumber\\
    &=-i\Tr \{\hat{Z}\Lcal_{kk'}\rho_k(0)\}\nonumber\\
    &=-i\Tr \{(\Lcal_{kk'}\rho_k(0))\hat{Z}\}\nonumber\\
    &=-\dfrac{\partial}{\partial k'}\Tr \{\Lcal_{kk'}\rho_k(0)\},\label{dk'}
\end{align}
Using \eqref{dk'} on the integration by parts of \eqref{jm_t} with $m=1$, we obtain
\begin{equation}
\braket{\hat{\mathbf{j}}}_t=-\int\dfrac{\mathrm{d}k}{2\pi}\sum_{n=1}^t
\Tr \{\hat{Z}\Lcal^n_{kk}\rho_k(0)\}.\label{j1_t}
\end{equation}
In the same way for \eqref{jm_t} with $m=2$,
\begin{equation}
\begin{split}
\braket{\hat{\mathbf{j}}^2}_t = -\int\dfrac{\mathrm{d}k}{2\pi}\left\{\sum_{n=1}^t\sum_{n'=1}^n \Tr \left[\hat{Z}\Lcal^{n-n'}_{kk} \left(\hat{Z}\Lcal_{kk}^{n'}\rho_k(0)\right)\right]\right.\qquad\\
+\left.\sum_{n=1}^t\sum_{n'=1}^{n-1} \Tr \left[\hat{Z}\Lcal^{n-n'}_{kk} \left((\Lcal_{kk}^{n'}\rho_k(0))\hat{Z}\right)\right]\right\}.\label{j2_t}
\end{split}
\end{equation}
It is possible to expand the states $\ket{\Phi_k(0)}$ in terms of the eigenstates of $ U_k$,
\begin{equation}
\ket{\Phi_k(0)}=c_k^+\ket{\Phi_k^+}+c_k^-\ket{\Phi_k^-},\label{phi_k0+-}
\end{equation}
in such a way that $c_k^{\pm}=\braket{\Phi_k^{\pm}|\Phi_k(0)}$. After inserting the \eqref{phi_k0+-} in \eqref{j1_t}, we obtain the expected value of position,
\begin{align}
\braket{\hat{\mathbf{j}}}_t &=-\sum_{n=1}^t\int_{-\pi}^{\pi}\dfrac{\mathrm{d}k}{2\pi}\bra{\Phi_k(0)}( U_k)^n\hat{Z}( U_k^\dagger)^n\ket{\Phi_k(0)}\nonumber\\
&=-t\int_{-\pi}^{\pi}\dfrac{\mathrm{d}k}{2\pi}\left\{ |c_k^+|^2\bra{\Phi_k^+}\hat{Z}\ket{\Phi_k^+}+ |c_k^-|^2\bra{\Phi_k^-}\hat{Z}\ket{\Phi_k^-} \right\}+\text{oscillatory terms},\label{j1}
\end{align}
and by means of \eqref{j2_t}, the expected value of the square of the position,
\begin{equation}
\begin{split}
\braket{\hat{\mathbf{j}}^2}_t =t^2\int_{-\pi}^{\pi}\dfrac{\mathrm{d}k}{2\pi}\left\{ |c_k^+|^2\bra{\Phi_k^+}\hat{Z}\ket{\Phi_k^+}^2+ |c_k^-|^2\bra{\Phi_k^-}\hat{Z}\ket{\Phi_k^-}^2 \right\}\\
+\mathcal{O}(t)+\text{oscillatory terms},\label{j2}
\end{split}
\end{equation}
where oscillatory terms vanish in the limit of $t\rightarrow\infty$. Therefore the expected values can be calculated as,
\begin{equation}
\braket{\Phi_k^{\pm}|\hat{Z}|\Phi_k^{\pm}}=\dfrac{\pm \cos(k-\delta)\left[ \sqrt{1+\cos^2(k-\delta)}\mp\cos(k-\delta) \right]}{1+\cos(k-\delta)\left[ \sqrt{1+\cos^2 (k-\delta)}\mp\cos(k-\delta)\right]},
\label{expect_value}
\end{equation}
and also the coefficients,
\begin{align}
c_k^{\pm}=\ &\dfrac{e^{-ik}}{N_k^{\pm}}\left\{\tilde{a}_k(0)-\tilde{b}_k(0)e^{i\eta}\left[  e^{i\delta}-\sqrt{2}\lambda_k^{\pm}e^{i(k-\delta)}\right]\right\},
\label{c_k}
\end{align}
where $\tilde{a}_k(0)=\tilde{f}(k)\cos(\alpha/2)$ and $\tilde{b}_k(0)=\tilde{f}(k)e^{i\beta}\sin(\alpha/2)$ are the spin up and down initial amplitudes in the $k$-space respectively. Therefore, to obtain the variance, we should insert these amplitudes in \eqref{c_k}, replacing them in \eqref{j1} together with \eqref{expect_value}. After these replacements, we have to solve the remaining integral,
\begin{equation}
I(\delta)=\int \dfrac{\mathrm{d}k}{2\pi}|\tilde{f}(k)|^2\left\{\dfrac{\cos^2(k-\delta)}{1+\cos^2(k-\delta)}\right\}
\label{Integral}
\end{equation}
to find
\begin{equation}
\braket{\hat{\mathbf{j}}}_t=I(\delta)\left[\cos\alpha+\sin\alpha\cos(\beta+\theta)\right]t ,
\label{j_general}
\end{equation}
for $t\gg 1$, since oscillatory terms are disregarded \cite{brun2003quantum}. This equation shows a dependence on the initial spin state ($\alpha$ and $\beta$) and coin parameter $\theta$. Nevertheless, the same does not occur with the square of the position, replacing \eqref{c_k} in \eqref{j2} with \eqref{expect_value} we reach,
\begin{equation}
\braket{\hat{\mathbf{j}}^2}_t=I(\delta)t^2.
\label{j2_general}
\end{equation}
Now we can insert both \eqref{j_general} and \eqref{j2_general} in \eqref{variance} to obtain the variance,
\begin{equation}
\sigma^2(t)=I(\delta)\left\{1-I(\delta)\left[\cos\alpha+\sin\alpha\cos(\beta+\theta)\right]^2\right\}t^2.
\label{var_general}
\end{equation}
In order to calculate the long-time variance for each initial state, we first should to write each particular $\tilde{f}(k)$, then replacing it in \eqref{Integral} and the result in \eqref{var_general} as shown on the next sections.

\subsection{Local state}

The local amplitudes in the $k-$space have $\tilde{f}_L(k)=1$ and inserting it in \eqref{Integral} gives $I_L=1-\sqrt{2}/2$. Then, replacing $I_L$ in \eqref{var_general}, we find the long-time variance, 
\begin{equation}
\sigma_L^2(t)=\left\{\left(1-\dfrac{\sqrt{2}}{2}\right)-\left(\dfrac{3}{2}-\sqrt{2}\right)\left[\cos\alpha+\sin\alpha\cos(\beta+\theta)\right]^2\right\}t^2,
\label{variance_Local}
\end{equation}
for an arbitrary initial spin state and coin. At this point, we are able to calculate the average variance by integrating over all spin states, 
\begin{equation}
\braket{\sigma_L^2}(t)=\int_{0}^{\pi}\dfrac{\mathrm{d}\alpha}{\pi}\int_{-\pi}^{\pi} \dfrac{\mathrm{d}\beta}{2\pi}\sigma_L^2(t)=\dfrac{2\sqrt{2}-1}{8}t^2,
\label{variance_mean_Local}
\end{equation}
where the dependence on the initial spin state ($\alpha$ and $\beta$) and coin parameter $\theta$ vanish.

\subsection{Gaussian state}

In order to obtain the Gaussian amplitudes in $k-$space, we should to change from the discrete variable $j$ to $x$ to integrate,
\begin{equation}
\tilde{f}_G(k)=\int\limits_{-\infty}^{+\infty}\frac{\text{exp}\left(-x^2/(4\sigma_0^2)-ikx \right)}{\left( 2\pi\sigma_0^2 \right)^{\frac{1}{4}}} \mathrm{d}x=\left(8\pi\sigma_0^2\right)^{\frac{1}{4}} e^{-k^2\sigma_0^2},
\label{f_Gauss}
\end{equation}
since imaginary part vanishes \cite{orthey2017asymptotic}. After replacing it in \eqref{Integral} the remaining integral does not have exact solution, however an approximate numerical solution results,
\begin{equation}
I_G(\delta,\sigma_0)=\dfrac{2\sigma_0}{\sqrt{2\pi}}\dfrac{\left(\mu\cos^4\delta+\nu\cos^2\delta+\xi\right)}{1+\cos^2\delta},
\label{I_Gauss}
\end{equation}
where $\mu,\nu$ and $\xi$ can be fitted by $\sum_{n=0}^4 a_n/\sigma_0^n$ whose parameters $a_n$ are in the Table~\ref{tab:1} of \ref{appendix_fit}. Inserting \eqref{I_Gauss} in \eqref{var_general}, we achieve a variance that is similar to the local case,
\begin{equation}
\sigma_G^2(t)=\{1-I_G(\delta,\sigma_0)\left[\cos\alpha+\sin\alpha\cos(\beta+\theta)\right]^2\}I_G(\delta,\sigma_0)t^2,
\label{variance_Gauss}
\end{equation}
except for the dependence on $\delta$ and $\sigma_0$, then the variance by averaging over all spin states gives,
\begin{equation}
\braket{\sigma_G^2}(t)=\left[1-\dfrac{3}{4}I_G(\delta,\sigma_0)\right]I_G(\delta,\sigma_0)t^2.
\label{variance_mean_Gauss}
\end{equation}
However, unlike the local case, we find a dependence on $\delta=(\theta+\phi)/2$ and $\sigma_0$ in \eqref{variance_mean_Gauss}. This result evidences that for an initial delocalized state, in particular a Gaussian one, the average variance still remains dependent on the quantum coin used along the walk and the initial dispersion of the state.

It is worth mentioning that the variance $\sigma_G^2(t)$ does not converge to $\sigma_L^2(t)$ for $\sigma_0\rightarrow 0$, in order to connect the Gaussian state to the local state (delta-like). Making it correctly would imply a renormalization of the state by means of a typical Error Function to maintain the condition of normalization \cite{orthey2017asymptotic}. However, the normalization of Gaussian states for $\sigma_0\geq1$ is preserved through this model.

\subsection{Uniform state}

At last, for a uniform state is easy to conclude that their $k-$space amplitudes have $\tilde{f}_G(k)=\sqrt{2\pi\delta(k)}$ and replacing it in \eqref{Integral} results,
\begin{equation}
I_U(\delta)=\dfrac{\cos^2\delta}{1+\cos^2\delta}.
\label{I_Uniform}
\end{equation}
Inserting \eqref{I_Uniform} in \eqref{var_general} gives us the variance,
\begin{equation}
\sigma_U^2(t)=\dfrac{\cos^2\delta+\cos^4\delta\{1-\left[\cos\alpha+\sin\alpha\cos(\beta+\theta)\right]^2\}}{(1+\cos^2\delta)^2}t^2,
\label{variance_Uniform}
\end{equation}
then a uniform state which evolves through a Fourier coin ($\theta=\phi=\pi/2)$ has a non-spreading behavior. The average variance is
\begin{equation}
\braket{\sigma_U^2}(t)=\dfrac{4\cos^2\delta+\cos^4\delta}{4(1+\cos^2\delta)^2}t^2,
\label{variance_mean_Uniform}
\end{equation}
which has a dependence on $\delta=(\theta+\phi)/2$ as the Gaussian case.

\subsection{Dispersion velocity}

\begin{figure}[h]
\center\includegraphics[width=0.8\linewidth]{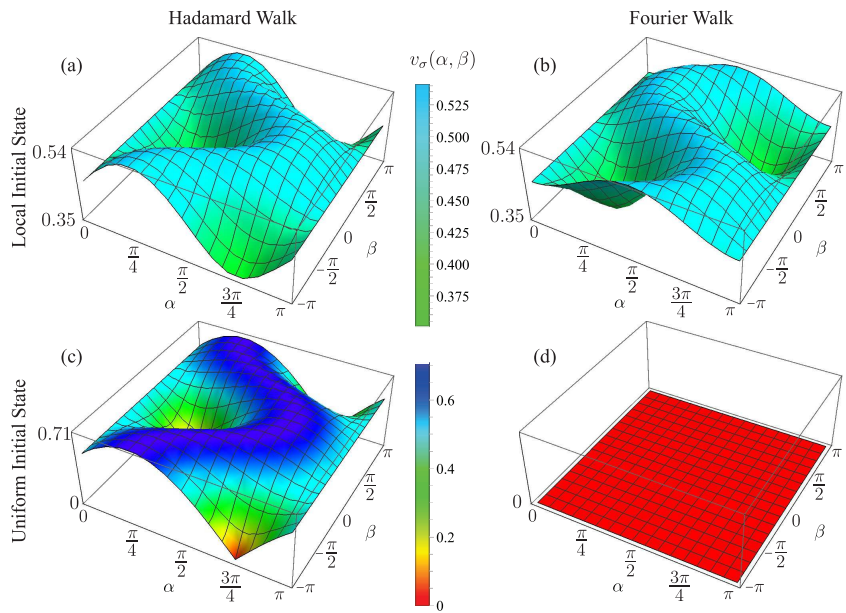}
\caption{Dispersion velocities $v_\sigma=\frac{\mathrm{d}\sigma(t)}{\mathrm{d}t}$ for Hadamard ($\theta=\phi=0$) and Fourier ($\theta=\phi=\pi/2$) walks starting from (a-b) local and (c-d) uniform states as function of initial spin states ($\alpha$ and $\beta$). All surfaces follow the same color scale. Purple regions have the highest values ($\sim 0.71$) and red ones have null values.}
\label{fig:1}
\end{figure}

The dispersion velocity of quantum walks starting from a local state driven by Hadamard and Fourier coins is distinguished only by a translation of $\pi/2$ in $\beta$ as shown in Fig. \ref{fig:1} (a) and (b).
This particular result corroborate the fact that there is no loss of generality on choosing any quantum coin when the quantum walks start from a local state \cite{tregenna2003controlling,ambainis2001one}. However, when they start from a uniform state by means of a Hadamard coin, they have a strong dependence on the initial spin state (qubit) and while by means of a Fourier coin, they are nondispersive regardless of the initial qubit as shown in Fig. \ref{fig:1} (c) and (d) respectively.

\section{Average spreading} \label{sec:4}

The quantum walks are very sensitive to their initial spin states, then in order to make a fair comparison between distinct position states and check the average quantities calculated via analytical approach, we carry out numerical calculations by averaging a large set of initial spin states. All averages are made over $N=2,016$ spin states varying $(\alpha,\beta)$ from $(0,0)$ to $(\pi,2\pi)$ in independent increments of $0.1$ of quantum walks starting from local state and a few Gaussian states. 
\begin{figure}[h]
\center\includegraphics[width=0.8\linewidth]{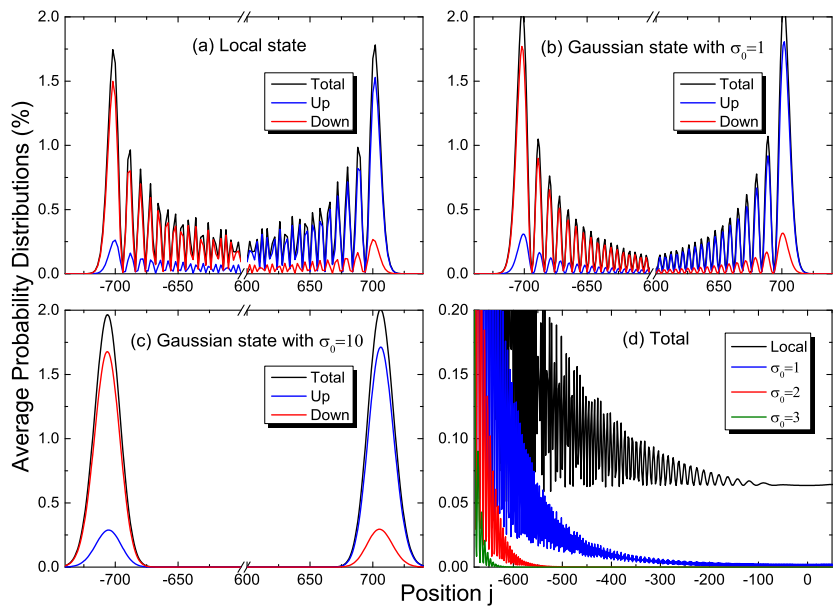}
\caption{Total average probability distributions (black) and for each spin component (red and blue) after $1000$ time steps of a Hadamard walk starting from (a) local and Gaussian states with initial dispersion (b) $\sigma_0=1$ and (c) $10$. For the sake of clarity, there is a break region between $j=-600$ and $600$. (d) A detail of the total average probability distributions for local (black) and Gaussian states with $\sigma_0=1$ (blue), $2$ (red) and $3$ (green). For the local case in (a) and (d), only the probabilities at the even points are plotted, once the odd points have null probability.}
\label{fig:2}
\end{figure}
Therefore, the total average probability distribution in each position $j$ at an arbitrary time $t$ is
\begin{equation}
\braket{|\Psi(j,t)|^2}=\sum_{i=1}^N\frac{|a_i(j,t)|^2}{N}+\sum_{i=1}^N\frac{|b_i(j,t)|^2}{N},
\label{Prob_Mean}
\end{equation}
where the terms on the right are the average probability distributions of spin up and down, respectively, and the index $i$ corresponding to each distinct initial spin state. In the same way, the average variance can be calculated by, 
\begin{equation}
\braket{\sigma^2}(t)=\sum_{i=1}^N\frac{\sigma^2_i(t)}{N}.
\label{sigma}
\end{equation}

\begin{figure}[h]
\center\includegraphics[width=0.8\linewidth]{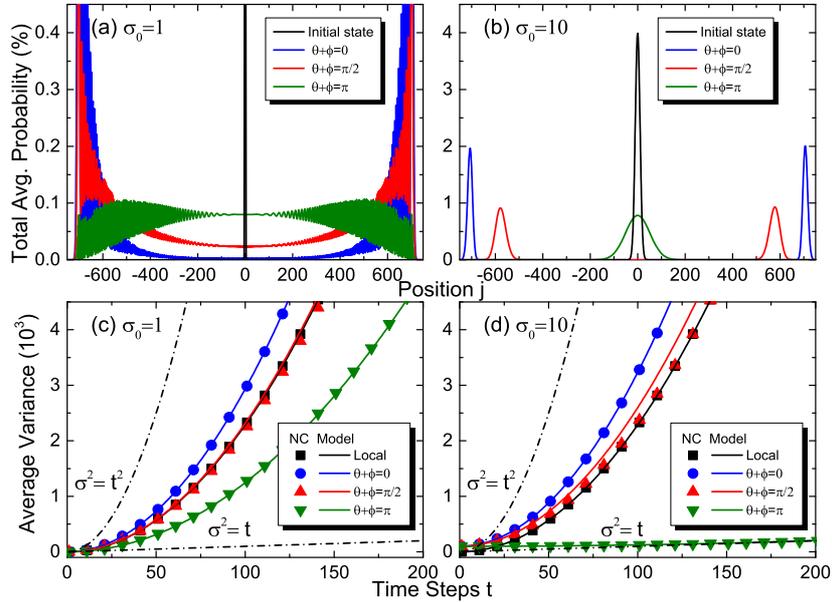}
\caption{Total average probability distributions starting from Gaussian states with initial dispersion (a) $\sigma_0=1$ and (b) $10$ with their initial states (black) at $t=0$ and after $1000$ time steps. These states time-evolve driven by quantum coins with $\theta+\phi=0$ (Hadamard in blue), $\theta+\phi=\pi/2$ (red) and $\theta+\phi=\pi$ (Fourier in green). Average variance of local state (black) and Gaussian states with (c) $\sigma_0=1$ and (d) $10$ obtained from numerical calculations (symbols) and from the expressions~\eqref{variance_mean_Local} and~\eqref{variance_mean_Gauss} (solid lines). Top and bottom dashed lines: $\sigma^2(t)=Ct^2$ and $\sigma^2(t)=Ct$ corresponding to ballistic and diffusive behavior with coefficient of proportionality $C=1$ as reference.}
\label{fig:3}
\end{figure}

We start our calculations using a Hadamard coin with $q=1/2$ and $\theta=\phi=0$, which creates a unbiased superposition between spin up and down without phase difference between them. Figure \ref{fig:2} shows the average probability distributions over the positions $j$ after $1000$ time steps of quantum walks starting from (a) local and (b)-(c) Gaussian states with initial dispersion $\sigma_0=1$ and $10$ respectively. For all cases, we obtain symmetrical total average distributions with both spin components in opposite sides. However, the spin up (down) has a greater probability on positive (negative) positions. The average probability ratio between spin up and down for positive or negative positions remains approximately steady along the time evolution and it decays asymptotically with the initial dispersion. For instance, let us consider $j<0$, then this ratio is about 33\% for local, 21\% and 18\% for Gaussian states with $\sigma_0=1$ and $2$ respectively and around 17\% from $\sigma_0=3$ and beyond. Figure \ref{fig:2} (d) shows a detail of the total average probability where we can see that, insofar as the Gaussian states delocalize ($\sigma_0$ increases), the probabilities decrease to zero far from the origin position, while the local state probability tends to a uniform distribution around $j=0$.

\begin{figure}[h]
\center\includegraphics[width=0.8\linewidth]{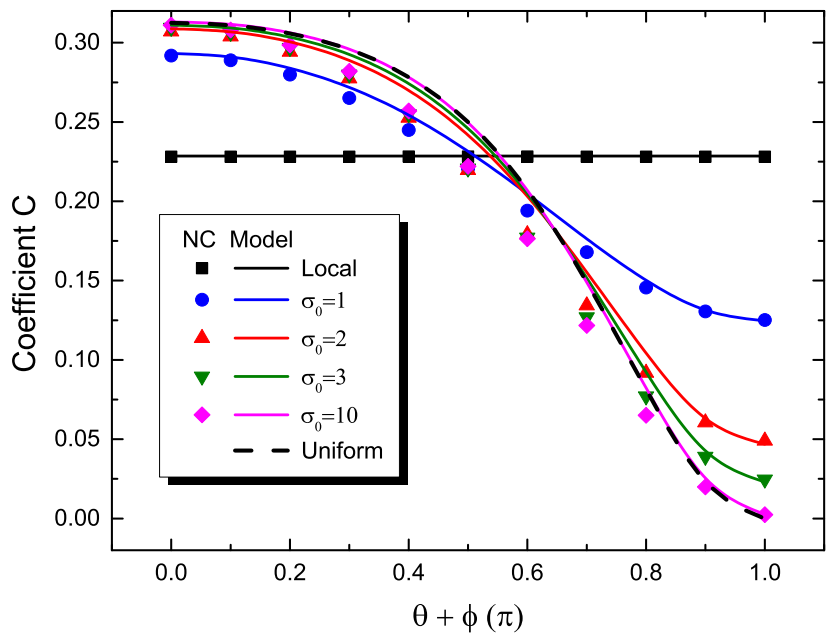}
\caption{Coefficient $C$ of $\braket{\sigma^2}=C t^2$. The symbols represent the coefficient $C$ extracted by fitting a polynomial function $f(t)=A+Bt+Ct^2$ of $\braket{\sigma^2}(t)$ obtained by means of numerical calculations (NC) of quantum walks after a $1000$ time steps for distinct values of $\theta+\phi$. Solid lines show the coefficients obtained from the expressions given by \eqref{variance_mean_Local} and \eqref{variance_mean_Gauss} for the local and Gaussian states, respectively. Dashed line corresponds to uniform state from \eqref{variance_mean_Uniform}. NC are for local state (black) and for Gaussian states with $\sigma_0=1$ (blue), $2$ (red), $3$ (green) and $10$ (pink).}
\label{fig:4}
\end{figure}

The average spreading behavior of quantum walks starting from a local state remains the same for all balanced coins ($q=1/2$) as showed in figure \ref{fig:2} (a). On the other hand, when the quantum walks starting from Gaussian states, they have a strong dependence on the parameters $\theta$ and $\phi$, peculiarly on the sum $\theta+\phi$ and also on the initial dispersion. In order to check the dependence between the initial dispersion and the quantum coin, we also carried out numerical simulations of quantum walks starting from Gaussian states driven by balanced coins from the Hadamard with $\theta+\phi=0$ up to the Fourier (or Kempe) coin with $\theta+\phi=\pi$ \cite{kempe2003quantum}. Figure \ref{fig:3} shows the total average probability of quantum walks starting from Gaussian states with initial dispersion (a) $\sigma_0=1$ and (b) $10$ and their average variances in (c) and (d) respectively. All cases display a quadratic behavior along the time, i.e., $\braket{\sigma^2}(t)=Ct^2$. However for a large initial dispersion, in particular for $\sigma_0=10$, the coefficient $C$ is very close to zero for $\theta+\phi=\pi$, which suggests a non-spreading behavior for $\sigma_0\gg 1$ in agreement with \eqref{variance_mean_Uniform}. Figure \ref{fig:4} shows how the coefficient $C$ varies for distinct values of $\theta+\phi$ and displays a comparison between a polynomial curve fit obtained from the simulations to their respective models for local and a few Gaussian states given by \eqref{variance_mean_Local} and \eqref{variance_mean_Gauss} respectively.

\section{Conclusions} \label{sec:5}

In this paper, we studied the spreading behavior of quantum walks through Brun-type formalism \cite{brun2003quantum} and numerical calculations. We obtained closed-form expressions for the long-time variance of position of quantum walks starting from any spin state (qubit) and local, Gaussian and uniform position states. We calculated the average variance analytically and we carried out extensive numerical calculations of the average variance and probability distribution profiles by averaging over a large ensemble of initial spin states. From both perspectives, we found out that the average variance of a quantum walk starting from a local state is always the same regardless the quantum coin, while from Gaussian and uniform states have a strong dependence on the quantum coin parameters, being non-dispersive for $\theta+\phi=\pi$ (Fourier walk) and $\sigma_0\gg 1$.

We hope our findings can be tested on different experimental platforms \cite{wang2013physical}. Particularly, it is important to notice that the external degree of freedom could be the $z$ component of orbital angular momentum instead of position $j$. In this context, the experiments based on the manipulation of the orbital angular momentum of photons from a unique light beam without refraction or reflections \cite{zhang2010implementation,goyal2013implementing,cardano2015quantum} seem to be promising for implementing delocalized states. Finally, it is worth mentioning that the resemblance between the asymptotic entanglement in quantum walks \cite{orthey2017asymptotic} and their long-time spreading behavior as function of their initial spin states suggests a relation between them, which might be a subject for a further study.

\section*{Acknowledgements}
ACO thanks CAPES (Brazilian Agency for the Improvement of Personnel of Higher Education) for the grant and Eduardo Luís Brugnago for useful discussions. EPMA thanks Janice Longo for her careful reading and considerations to improve the manuscript.

\appendix

\section{Fitted parameters for the variance of Gaussian states}\label{appendix_fit}

The values $\mu$, $\nu$ and $\xi$ from the function $I_G(\delta,\sigma_0)$ in \eqref{I_Gauss} following the model $\sum_{n=0}^{4}a_n/\sigma_0^n$, whose parameters $a_n$ are in the Table \ref{tab:1}.
\begin{center}
\begin{table}[ht]
\center\begin{tabular}{lccc} 
\hline
     & $\mu$              & $\nu$              & $\xi$ \\
\hline 
$a_0$  &  0.0022$\pm$0.0004 & -0.0020$\pm$0.0005 &  0.0002$\pm$0.0001 \\
$a_1$  & -0.0492$\pm$0.0077 &  1.2995$\pm$0.0085 & -0.0053$\pm$0.0020 \\ 
$a_2$  &  0.2938$\pm$0.0361 & -0.2668$\pm$0.0400 &  0.0296$\pm$0.0095 \\
$a_3$  &  0.5030$\pm$0.0596 & -1.0016$\pm$0.0661 &  0.2548$\pm$0.0157 \\
$a_4$  & -0.4612$\pm$0.0312 &  0.5991$\pm$0.0346 & -0.1049$\pm$0.0082 \\
\hline
\end{tabular}
\caption{Fitted parameters and standard errors of $\mu$, $\nu$ and $\xi$.}\label{tab:1}
\end{table}
\end{center}

\end{document}